\documentclass[paper]{elsarticle}

\usepackage{amsfonts,amsbsy,amssymb,amsmath}
\usepackage{lineno,hyperref}
\usepackage{graphicx}
\usepackage[table,xcdraw]{xcolor}
\usepackage{float}

\graphicspath{{figures/}}

\modulolinenumbers[5]

\biboptions{compress}

\begin{document}

\begin{frontmatter}

\title{The bifurcations of the critical points and the role of the depth in a symmetric Caldera potential energy surface}


 \author[label1]{Y. Geng}
 \author[label1]{M. Katsanikas}
 \author[label1]{M. Agaoglou}
 \author[label1]{S. Wiggins\corref{mycorrespondingauthor}}
 \ead{s.wiggins@bristol.ac.uk}

  \address[label1]{School of Mathematics, University of Bristol, \\ Fry Building, Woodland Road, Bristol, BS8 1UG, United Kingdom.\\[.2cm]}

\begin{abstract}
In this work, we continue the study of the bifurcations of the critical points in a symmetric Caldera potential energy surface. In particular, we study the influence of the depth of the potential on the trajectory behavior before and after the bifurcations of the critical points. We observe two different types of trajectory behavior:  dynamical matching and the non-existence of dynamical matching. Dynamical matching is a phenomenon that limits the way in which a trajectory can exit the Caldera based solely on how it enters the Caldera. Furthermore, we discuss  two different types of symmetric Caldera potential energy surface and the transition from the one type to the other through the bifurcations of the critical points.
\end{abstract}


\end{frontmatter}


\section{Introduction}
\label{intro}
In this paper we study the  influence of the depth of the potential on the trajectory behavior before and after  a bifurcation of critical points in a symmetric Caldera potential energy surface (PES). Doering \cite{doering1968} gave this name to this potential energy surface because its morphology  is similar to this of a collapsed region in an erupted volcano. The Caldera PES occurs in  many organic chemical reactions,  such as  the   vinylcyclopropane-cyclopentene rearrangement \cite{baldwin2003,gold1988}, the stereomutation of cyclopropane \cite{doubleday1997}, the degenerate rearrangement of bicyclo[3.1.0]hex-2-ene \cite{doubleday1999,doubleday2006} or that of 5-methylenebicyclo[2.1.0]pentane \cite{reyes2002}.

We have two types  of symmetric  Caldera potential energy surfaces. The transition between the two types of potential energy surfaces occurs as a result  of a pitchfork bifurcation of the critical points. The first type is characterized  by a shallow minimum that is surrounded by potential walls and four index-1 saddles that exist around these potential walls (see \cite{collins2014,katsanikas2018} and panel B of Fig. \ref{3Dcontours}). In this case, the chemical reaction takes place when the trajectories that have initial conditions in the region of the upper index-1 saddles  cross the Caldera and  approach one of the two lower index-1 saddles. Then the trajectories leave the Caldera through the region of one of the lower index-1 saddles.

 The second type is characterized by three index-1 saddles, two upper index-1 saddle and one lower index-1 saddle that is on the $y$-axis (see \cite{geng2021influence} and panels A and C of Fig. \ref{3Dcontours}). In this case the chemical reaction takes place when  trajectories that have initial conditions in the region of the upper index-1 saddles  cross the Caldera and they approach the lower index-1 saddle. Then the trajectories  leave the Caldera through one of the regions  that exist on both sides of the lower index-1 saddle. We will refer to these regions as the lower right exit region and the lower left exit region. 

We have two categories of trajectory behavior in the symmetric Caldera potential energy surface:

\begin{enumerate}
    \item {\bf Dynamical matching:} This phenomenon was first reported in \cite{carpenter1985,carpenter1995}. In this case the trajectories (reactants) that begin from the  region of the upper index-1 saddle cross the Caldera. Then they exit  through the region of the opposite lower saddle (in the case of the first type of  symmetric Caldera potential) or through the region of the opposite lower exit region   (for the case of the second type of  symmetric Caldera potential). This means that  $100\%$ of the trajectories are exiting from one of the two  lower exits and not $50\%$, as  would be predicted by statistical theories.
    
    \item {\bf Non-existence of dynamical matching:}  This category includes the trajectories that begin from the region of the upper index-1 saddles and they  do not obey dynamical matching. This means that the trajectories  are trapped in the intermediate region of the Caldera before they exit. In addition, the trajectories can exit through either region of lower saddles (in the case of the first type of symmetric Caldera potential) or any lower exit region (in the case of the second type of symmetric Caldera potential).
\end{enumerate}

 The mechanism that controls the existence of  the dynamical matching is the  existence  of heteroclinic intersections  between the unstable manifolds of the unstable periodic orbits associated with the upper saddles and the stable manifolds of invariant sets in the central region. If we do not have the existence of these heteroclinic intersections, we encounter the phenomenon of dynamical matching (see \cite{katsanikas2018}) otherwise  we don't have this phenomenon (see \cite{katsanikas2020a,katsanikas2020b,katsanikas2019}).

In this paper, we want to study the influence of the depth of the potential (a geometrical feature of the potential - see the definition in Section \ref{results}) on the trajectory behavior in  the symmetric Caldera potential energy surface. In the previous papers  (\cite{collins2014,katsanikas2018,geng2021influence}) we saw that  dynamical matching occurs in the first type of the symmetric Caldera potential energy surface and the non-existence of dynamical matching in the second type of the symmetric Caldera potential energy surface. This change of the trajectory behavior is abrupt and it  happens just after the transition  from  one type of the symmetric Caldera potential energy surface to the other (this transition is a result of a pitchfork bifurcation of critical points -see \cite{geng2021influence}). This transition that is studied in \cite{geng2021influence} happens at a small value of the depth of the potential  (see section \ref{results}). In this paper we want to investigate  the trajectory behavior when this transition occurs at large values of the depth of the potential. In this way, we want to determine if the existence or the non-existence of dynamical matching is a general characteristic of the first and second type of  symmetric Caldera potential energy surface respectively. Furthermore, we want to study if the change of the trajectory behavior is abrupt or not in the cases that the transition between the two kinds of the symmetric Caldera  potential energy surface occurs for large values of the depth of the potential.

In this paper is outlined as follows. In Section \ref{model} we present the Hamiltonian model and the potential of our system (section \ref{model}). In Section \ref{results} we describe the trajectory behavior before and after the first and the second bifurcation and we compare the topography
of the PES before and after the bifurcations. Finally in Section \ref{conclusions} we summarize our findings.

\section{Model}
\label{model}

In this section we introduce the model of the symmetric, with respect to the y-axis, Caldera potential energy surface (PES). The expression of the PES that has been introduced in \cite{collins2014} for  2 degree of freedom (DoF) Hamiltonian systems is given by:

\begin{equation}
    \begin{aligned}
    V(x,y) &= c_1r^2+c_2y-c_3r^4\cos(4\theta)\\ &= c_1(x^2+y^2)+c_2y-c_3(x^4+y^4-6x^2y^2),
    \end{aligned}
\end{equation}

\noindent
where $(x,y)$, $(r,\theta)$ describe the position in Cartesian and polar coordinates, respectively, and $c_1,c_2,c_3$ are  parameters.

The 2 DoF Hamiltonian is given by:

\begin{equation}
    H(x,y,p_x,p_y)=\frac{p_x^2}{2m}+\frac{p_y^2}{2m}+V(x,y),
\end{equation}

\noindent
where $p_x$ denotes the momentum at $x$ and $p_y$ denotes the momentum at $y$ and $m$ is a considered to be $1$. Thus the equations of motion are:

\begin{equation}
    \begin{aligned}
    &\dot{x} = \frac{\partial H}{\partial p_x} = \frac{p_x}{m}\\
    &\dot{y} = \frac{\partial H}{\partial p_y} = \frac{p_y}{m}\\
    &\dot{p_x} = - \frac{\partial V}{\partial x}(x,y) = -(2c_1x-4c_3x^3+12c_3xy^2)\\
    &\dot{p_y} = - \frac{\partial V}{\partial y}(x,y) = -(2c_1y-4c_3y^3+12c_3x^2y+c_2)
    \end{aligned}
\end{equation}

\section{Results}
\label{results} 

In this section, we will present our results. We will study the trajectory behavior before and after two  pitchfork bifurcations of critical points. As we  described in the Section \ref{intro}, these bifurcations represent a transition from one kind of the  symmetric Caldera potential energy surface to the other. The first bifurcation occurs  at a small value of the depth of the potential (we define this in subsection \ref{depth0})   and the second 
bifurcation at a large value. 

In order to find the first bifurcation (a bifurcation of critical points that occurs at a small value of the depth of the potential) we fixed the potential parameters $c_2=3$ and $c_3=-0.3$ (that are used in previous papers see for example \cite{collins2014,katsanikas2018} ) and we vary the parameter $c_1$ between $0$ and $5$. As a consequence, we have a bifurcation of critical points at $c_1=1.32$ (see more details at the subsection \ref{bif0}) that  happens at a small value of the depth of the potential (6.228 - see Fig. \ref{depthdef1}). In order to find the second bifurcation (a bifurcation of critical points that occurs at a large value of the depth of the potential) we fixed the potential parameters $c_1=5$ and $c_3=-0.3$ (that are used in previous papers see for example \cite{collins2014,katsanikas2018} ) and we vary the parameter $c_2$ between $0$ and $162$. This results a bifurcation of critical points at $c_2=22.22$ (see more details at the subsection \ref{bif0}) that  happens at a large value of the depth of the potential (89.77 - see Fig. \ref{depthdef1}). This means that fixing the parameters $c_2$ and $c_3$ and varying $c_1$, we find bifurcations of critical points that occur at small values of the depth of the potential. On the contrary, if we fix the parameters $c_1$ and $c_3$ and vary $c_2$, we find bifurcations of critical points that occur at large values of the depth of the potential.

In this section, we firstly give the definition of the depth of the potential (see the Subsection \ref{depth0}). Secondly, we describe the first and second bifurcation of critical points that occur at a small and a large value of the depth of the potential respectively (see the Subsection \ref{bif0}). Finally,  we describe the trajectory behavior  before and after the first and the second bifurcation of critical points in the Subsections \ref{bif1} and \ref{bif2} respectively.

\subsection{Depth of the potential}
\label{depth0}
In this  section, we give the definition of the depth of the potential 
for the first kind and the second kind of the symmetric Caldera potential energy surface. As we described above (Section \ref{intro}) the first kind corresponds to a  symmetric Caldera potential energy surface with one center and four index-1 saddles  around it (two for high values of energy and two for low values of energy). The second  kind corresponds to a  symmetric Caldera potential energy surface with three index-1 saddles (two for high values of energy and one for low values of energy).

We define the depth of the potential as the difference between the potential energy that corresponds to the high energy saddles and  the potential energy of the center at the bottom of the well (for the case of the first kind of the symmetric Caldera potential energy surface): 

\begin{equation}
D=V(x_{highersaddle},y_{highersaddle})-V(x_{center},y_{center})
\label{depth1}
\end{equation}

\noindent
where $x_{highersaddle},y_{highersaddle}$  and $x_{center},y_{center}$ are the $x$ and $y$ coordinates in the configuration space  of the high energy saddle (one of them) and of the center respectively. In the case that we have a lower index-1 saddle  instead the center at the bottom (for the case of the second kind of the symmetric Caldera potential energy surface), the depth of the potential is defined  as the difference between  the potential energy that corresponds to the high energy saddles and the potential energy of the lower energy saddle:

\begin{equation}
D=V(x_{highersaddle},y_{highersaddle})-V(x_{lowersaddle},y_{lowersaddle})
\label{depth2}
\end{equation}

\noindent
where $x_{highersaddle},y_{highersaddle}$  and $x_{lowersaddle},y_{lowersaddle}$ are the $x$ and $y$ coordinates in the configuration space  of the high energy saddle (one of them) and of the lower energy saddle respectively. In Fig. \ref{depthdef1} we can see how the depth of the potential is changing while we modify the parameter $c_{1}$ and $c_2$.

\subsection{Bifurcations of critical points}
\label{bif0}

\begin{figure}
 \centering
A)\includegraphics[scale=0.4]{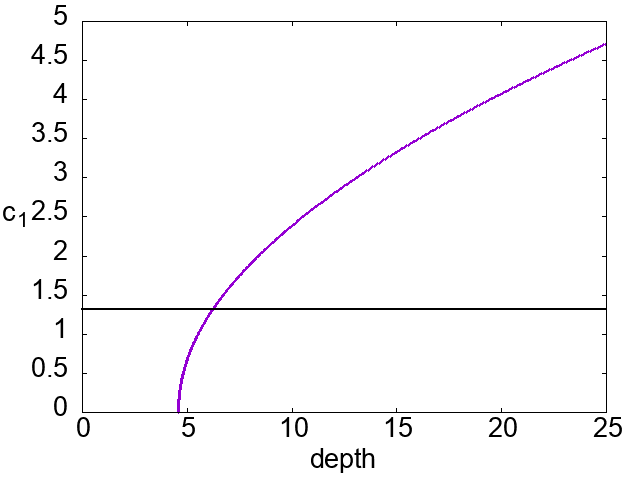}
B)\includegraphics[scale=0.4]{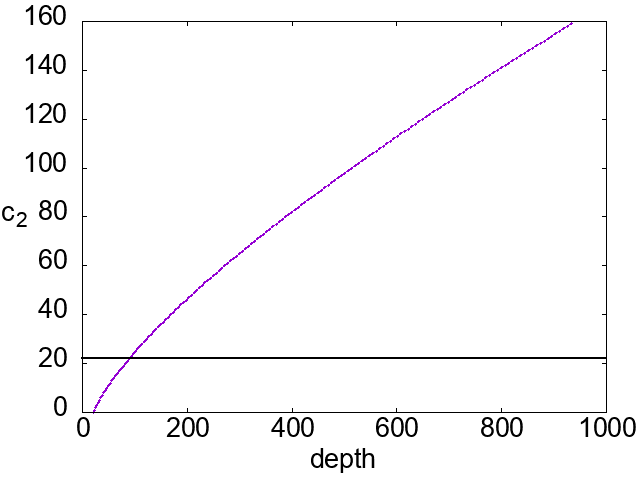}\\
\caption{A) The evolution  of the $c_1$ versus the depth of the potential. The black line indicates the value $c_1=1.32$ (for which a bifurcation occurs) and it corresponds to a value 6.228 for the depth of the potential. B) The evolution  of the $c_2$ versus the depth of the potential. The black line indicates the value  $c_2=22.22$ (for which a bifurcation occurs) and it corresponds to a value 89.77 for the depth of the potential.    }
\label{depthdef1}
\end{figure}

In this section we discuss two  bifurcations of critical points that occur on the Caldera PES. The first bifurcation occurs at a small value of the depth of the potential (6.228 - see Fig. \ref{depthdef1}) and the second bifurcation occurs at a large value of the depth of the potential (89.77 - see Fig. \ref{depthdef1}).

\begin{figure}
 \centering
A)\includegraphics[scale=0.15]{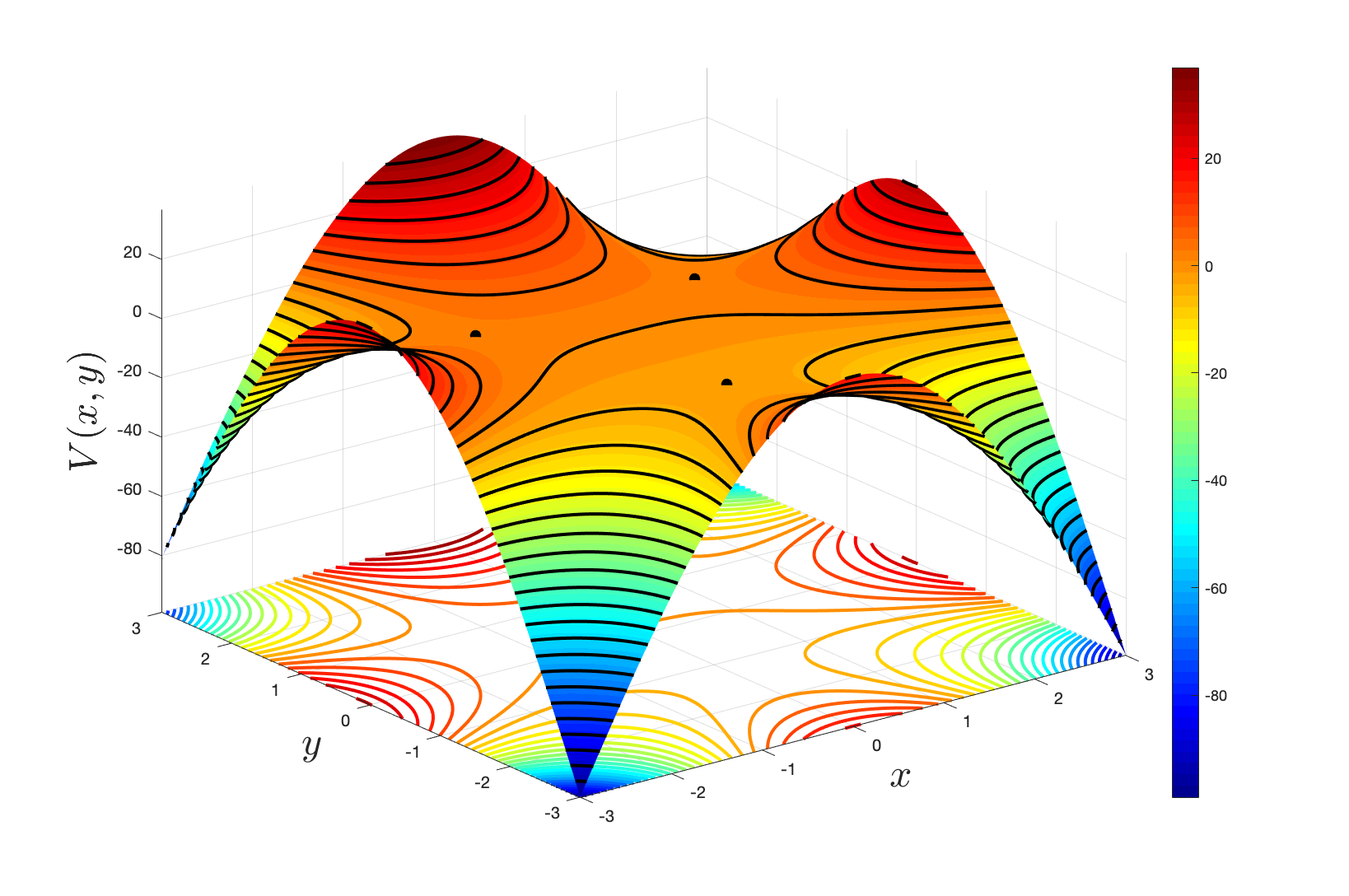}
B)\includegraphics[scale=0.15]{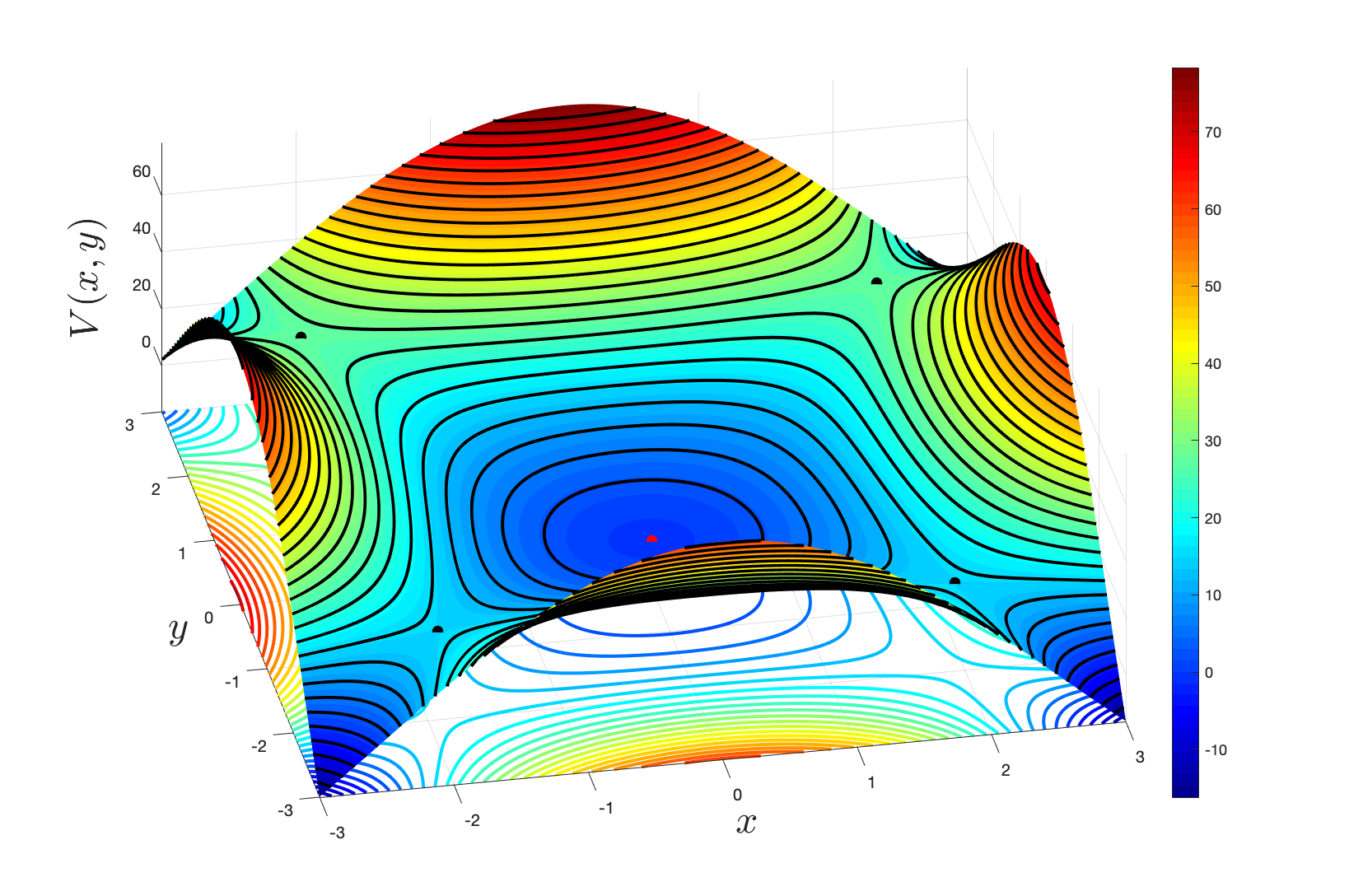}\\
C)\includegraphics[scale=0.15]{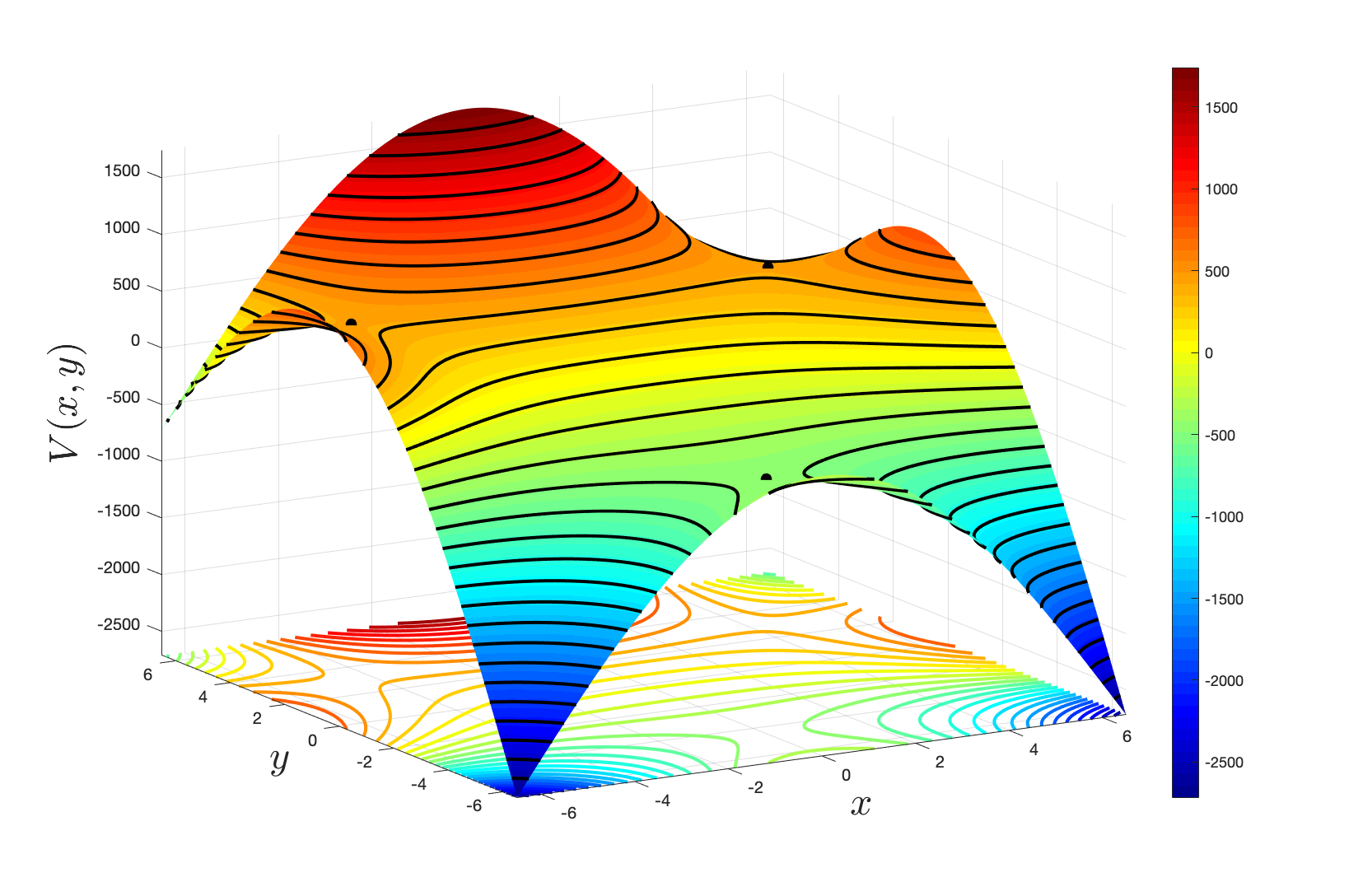}\\
\caption{The 3D PES of the Caldera potential A) for  $c_1=0.4,c_2=3$ and $c_3=-0.3$ B) for $c_1=5,c_2=3$ and $c_3=-0.3$ and C) for for $c_1=5,c_2=153$ and $c_3=-0.3$. The index-1 saddles and the center are depicted with black and red points, respectively.}
\label{3Dcontours}
\end{figure}

\begin{figure}
 \centering
A)\includegraphics[scale=0.3]{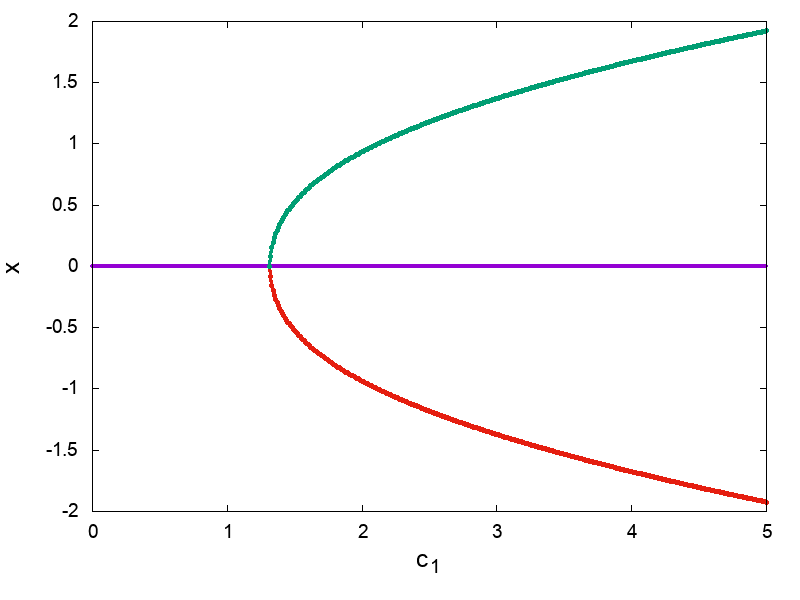}
B)\includegraphics[scale=0.3]{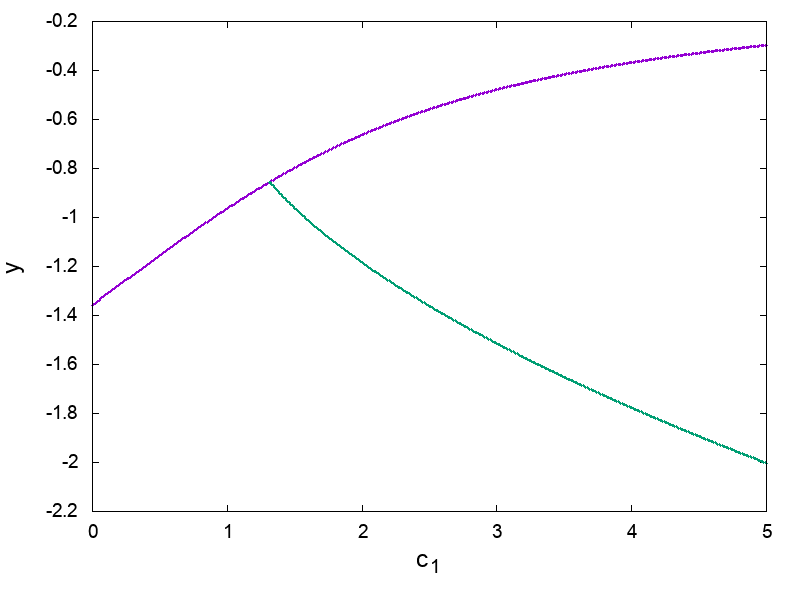}\\
\caption{The first bifurcation diagrams of lower region of the Caldera potential: A) with respect to $x$-coordinate. B)  with respect to $y$-coordinate.}
\label{bifurcations1}
\end{figure}

In \cite{geng2021influence} the potential parameters $c_2=3$ and $c_3=-0.3$ are fixed while the parameter $c_{1}$ is varied between $0$ and $5$. In this situation we show that there is  a pitchfork bifurcation of the critical points of the Caldera PES. In particular, for $0 \le c_1 \le 1.32$, the Caldera PES has three critical points in total, two symmetric, with respect to y-axis, index-1 saddles in the upper plane (upper LH and upper RH saddles) and one index-1 saddle on the negative $y$-axis (lower saddle), see the panel A of Fig. \ref{3Dcontours}. After the critical value $c_1=1.32$, the lower saddle bifurcates into two symmetric index-1 saddles in the lower plane and one center on the $y$-axis. Therefore the Caldera PES after the critical value $c_1 = 1.32$ has five critical points, see panel B of Fig. \ref{3Dcontours}. The bifurcation diagrams are shown in Fig. \ref{bifurcations1}

\begin{figure}
 \centering
A)\includegraphics[scale=0.55]{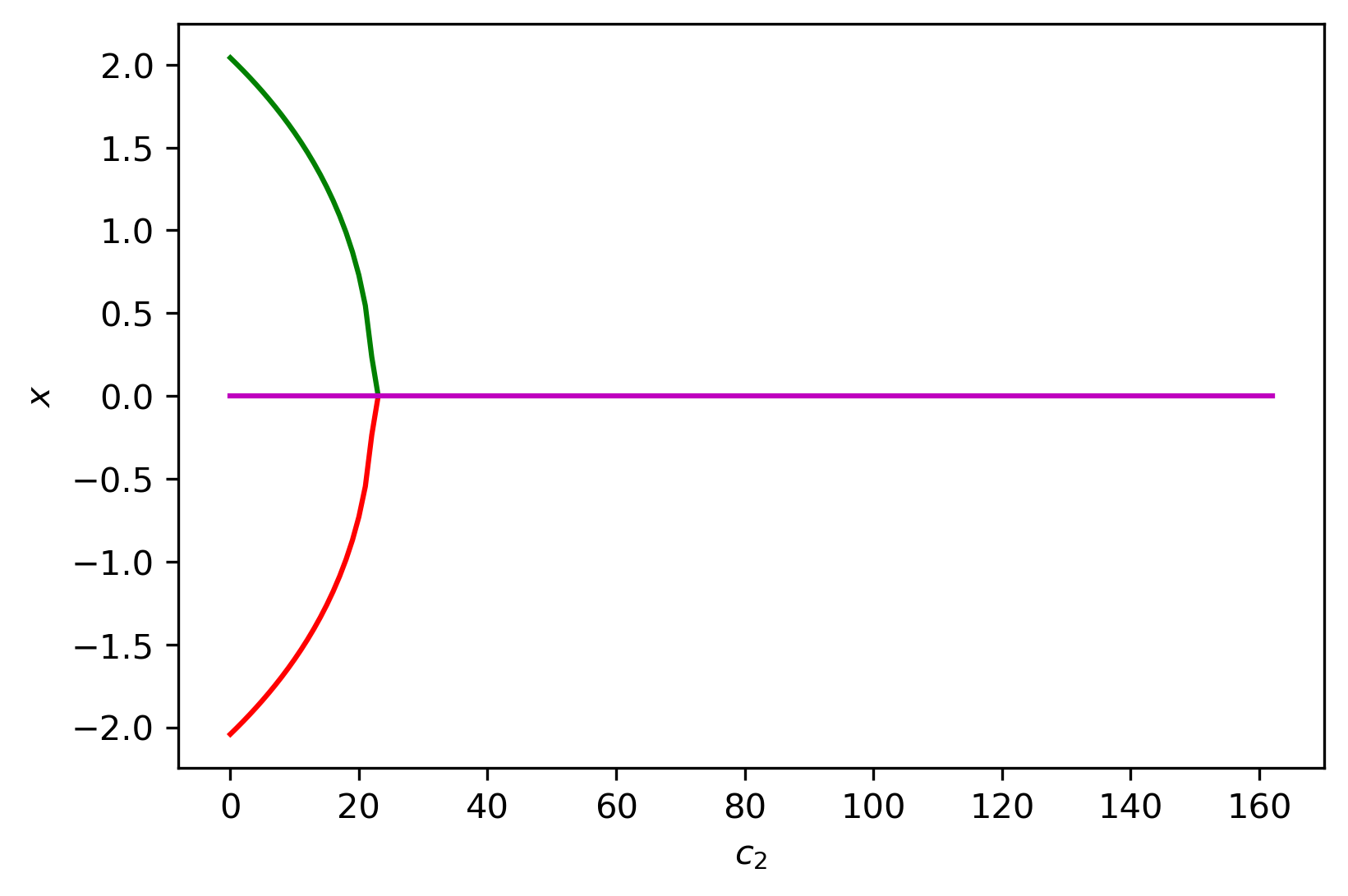}
B)\includegraphics[scale=0.55]{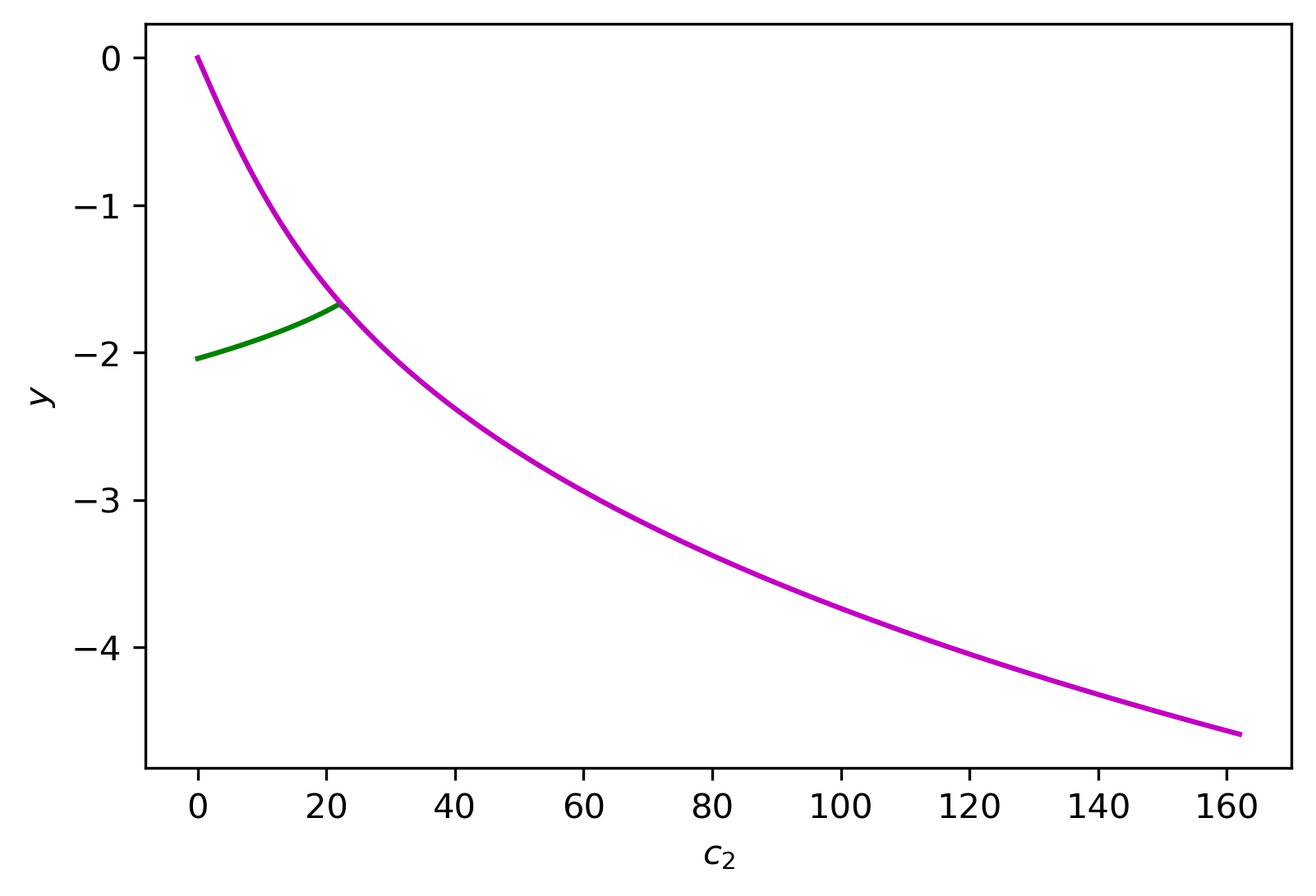}\\
\caption{The second bifurcation diagrams of lower region of the Caldera potential: A) with respect to the $x$-coordinate. B)  with respect to the $y$-coordinate.}
\label{bifurcations2}
\end{figure}

The second bifurcation occurs when we fix $c_1 = 5,c_3 = -0.3$ and vary the $c_2$ parameter. This also gives us a pitchfork bifurcation. When $c_2 \le 22.22$ the system has five critical points i.e. a centre (which is a minimum) and four index-1 saddles around it. A typical example is the case when $c_2=3$, the four index-1 saddles control the exit from and the entrance to the Caldera, see  panel B of Fig. \ref{3Dcontours}. After the critical value of $c_2=22.22$ the number of critical points changes from five to three, see panel C of Fig. \ref{3Dcontours}, where the two lower saddles and the centre coincide at a lower index-1 saddle. In this case, while the upper critical points still control the entrance to and exit from the region of their directions, the single lower critical point controls the entire lower region, both lower left and lower right. We will refer to these region as the lower left exit region and the lower right exit region. The bifurcation diagrams  are given in Fig. \ref{bifurcations2}

The position of all critical points for the parameter values that we chose in every panel in Fig. \ref{3Dcontours}, and their energies are given in Table \ref{table:1} and Table \ref{table:2}.

\begin{table}
\begin{center}
\caption{Stationary points of the Caldera potential for $c_1=0.4,c_2=3$ and $c_3=-0.3$  ("RH" and "LH" are the abbreviations for right hand and left hand respectively)}
\label{table:1}
\begin{tabular}{l  l  l  l}
\hline
Critical point & x & y & E \\
\hline
Lower  saddle & 0.000 & -1.194 & -2.402 \\
Upper LH saddle  & -1.204 & 0.840  & 2.321 \\
Upper RH saddle  & 1.204 &  0.840 & 2.321 \\
\hline
\end{tabular}
\end{center}
\end{table} 

\begin{table}
\begin{center}
\caption{Stationary points of the Caldera potential for $c_1=5,c_2=3$ and $c_3=-0.3$  ("RH" and "LH" are the abbreviations for right hand and left hand respectively)}
\label{table:2}
\begin{tabular}{l  l  l  l}
\hline
Critical point & x & y & E \\
\hline
Central minimum & 0.000 & -0.297 & -0.448 \\
Upper LH saddle  & -2.149 & 2.0778 & 27.0123 \\
Upper RH saddle  & 2.149 &  2.0778 & 27.0123 \\
Lower LH saddle & -1.923 & -2.003  & 14.767 \\
Lower RH saddle & 1.923 & -2.003 & 14.767 \\
\hline
\end{tabular}
\end{center}
\end{table}

\begin{table}
\begin{center}
\caption{Stationary points of the Caldera potential for $c_1=5,c_2=153$ and $c_3=-0.3$  ("RH" and "LH" are the abbreviations for right hand and left hand respectively)}
\label{table:3}
\begin{tabular}{l  l  l  l}
\hline
Critical point & x & y & E \\
\hline
Lower  saddle & 0.000 & -4.484 & -464.242 \\
Upper LH saddle  & -4.448 & 3.061  & 424.194 \\
Upper RH saddle  & 4.448 &  3.061 & 424.194 \\
\hline
\end{tabular}
\end{center}
\end{table}

\subsection{Trajectory behavior before and after the first Bifurcation}
\label{bif1}

In this section we discuss the behaviour of trajectories initialised in the upper region for $0\le c_1\le5$ with $c_2=3$ and $c_3=-0.3$, and determine whether they exit the Caldera through the region of either saddle or are trapped in the Caldera.

First, we will describe how the initial conditions of trajectories were chosen. Since the Caldera PES is symmetric, without loss of generality we only constructed trajectories initialised in the upper right region. While the initial conditions consist of the position and the momentum, we chose $1000$ initial positions on the line which passes through the upper right index-1 saddle, perpendicular to the line joining the upper right and lower left index-1 saddle for $1.32\le c_1\le 5$ or the line joining the upper right index-1 saddle and the lower saddle on the $y$-axis for $0\le c_1\le 1.32$. The momentum at that position is in the direction of the line which connects two index-1 saddles, and the energy is constant throughout the evolution of the trajectory. The $1000$ initial positions were uniformly distributed along the line segment.

We integrated all the initial conditions for a fixed time $t=3$ (we used the same value as in \cite{geng2021influence}) and determined the region where they were located after this time interval. In particular,  we counted the number of trajectories that passed through each exit region  or that became trapped  in the intermediate region of the Caldera.  Then we divided these numbers by 1000, the total number of trajectories, in order to find the ratio of the trajectories for every exit region and the ratio of the trajectories that were trapped.

\begin{figure}
 \centering
A)\includegraphics[scale=0.55]{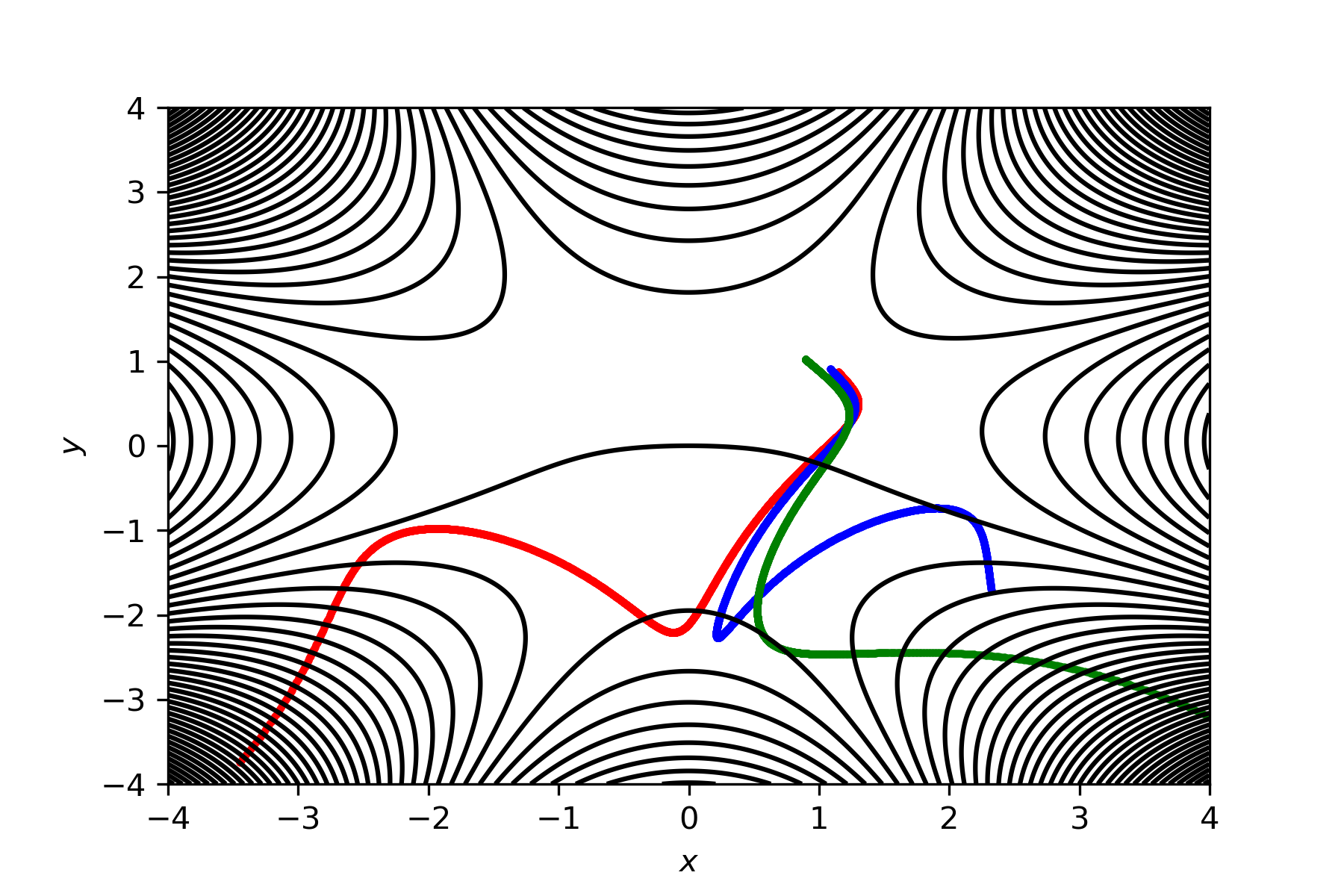}
B)\includegraphics[scale=0.62]{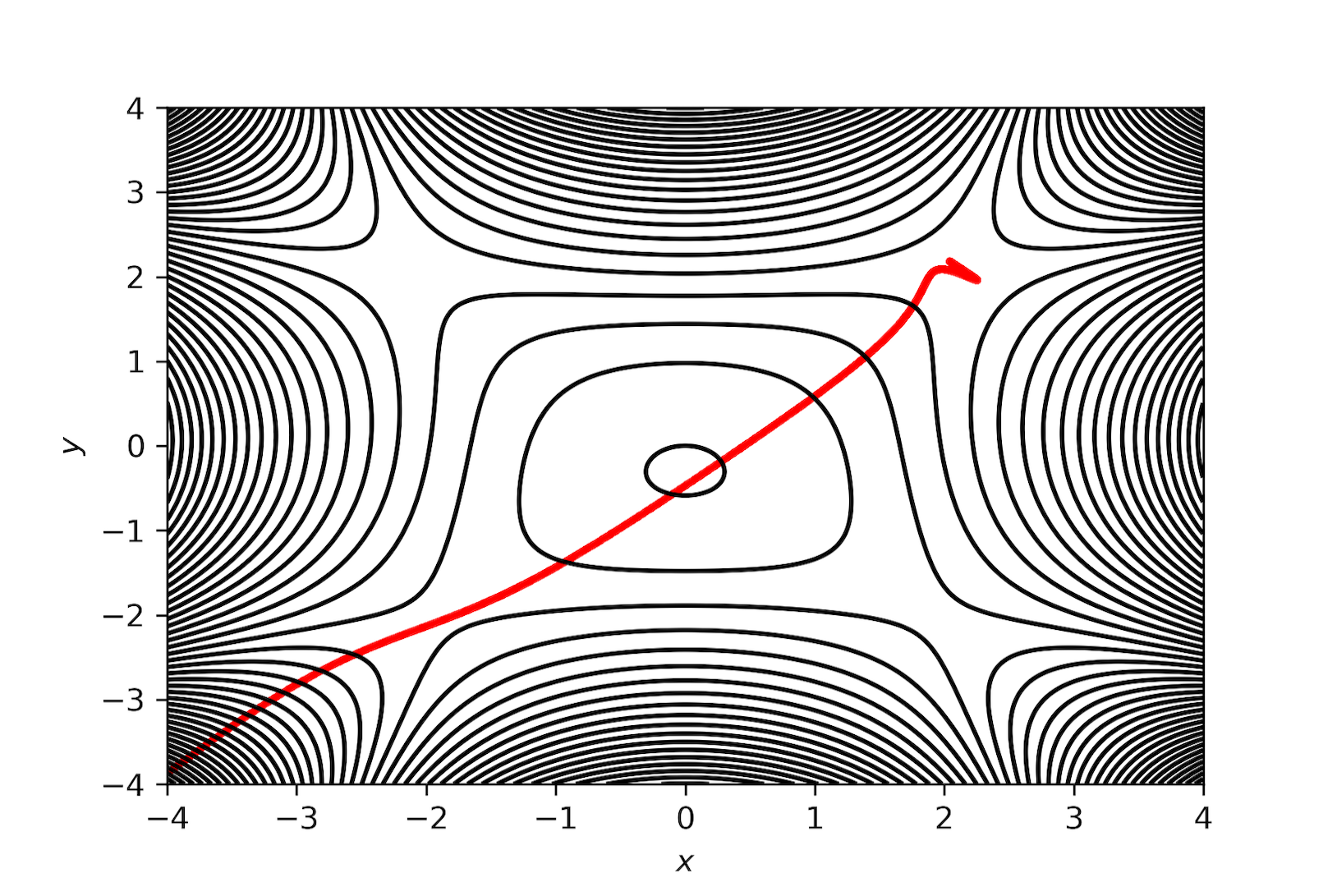}
C)\includegraphics[scale=0.55]{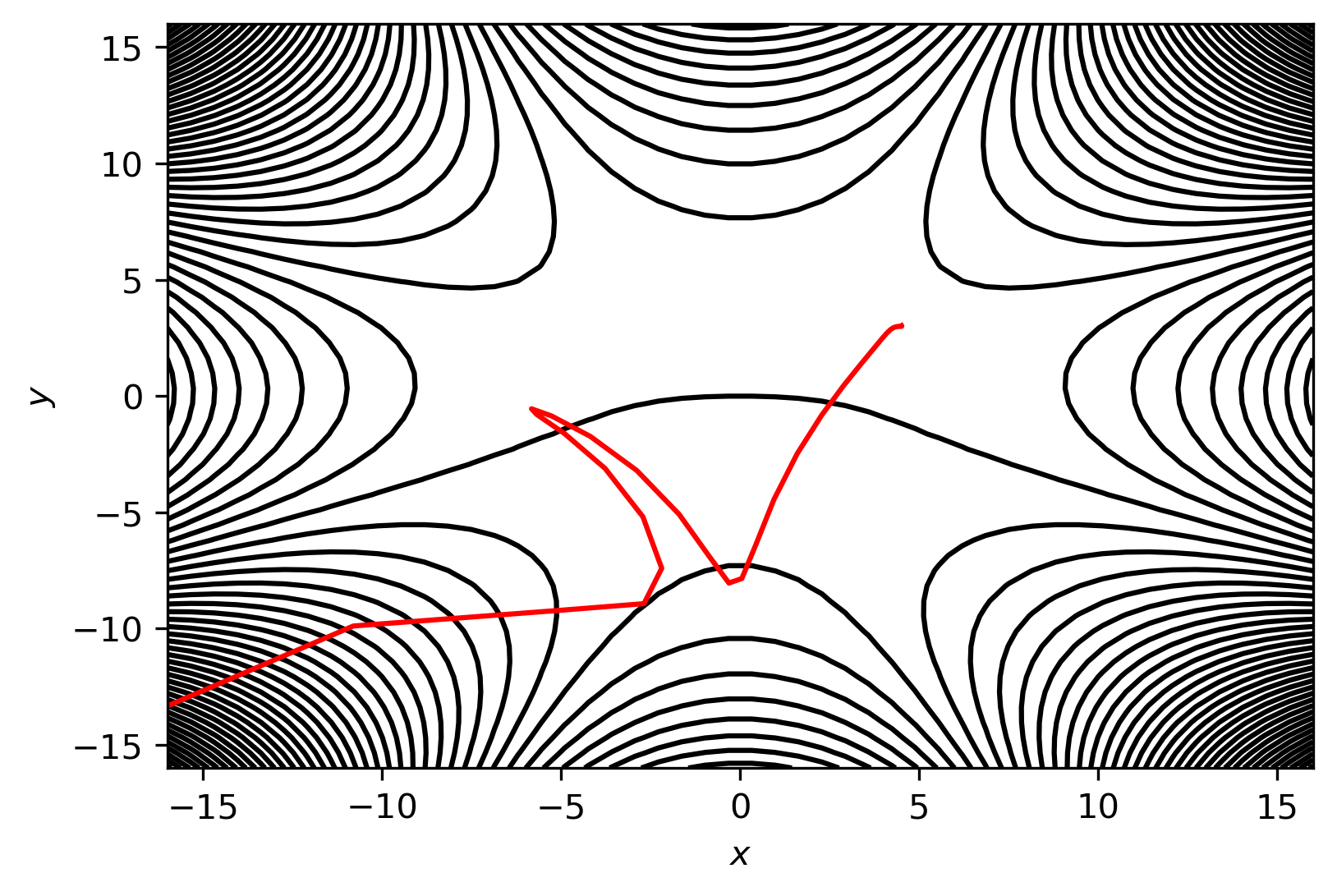}\\
\caption{The contours of the Caldera potential and  trajectories in the configuration space that begin from the region of the upper right saddle (for a value of energy 0.5 units above the energy of the higher index-1 saddles): A) for $c_1=0.4,c_2=3$ and $c_3=-0.3$, B)  for $c_1=5,c_2=3$ and $c_3=-0.3$ (The panels A and B are adapted from \cite{geng2021influence}), C)  for $c_1=5,c_2=153$ and $c_3=-0.3$. The trajectories that are trapped or that exit through the region of the lower left saddles (or the lower left exit region)  or that exit through the region of the lower right saddle (or the lower right exit region) are depicted by blue, red and green colors, respectively.}
\label{traj}
\end{figure}

From these results we observed that for $0 \le c_1\le1.32$ the trajectories  enter the lower left exit region, the lower right exit region or are trapped in the Caldera. In panel A of Fig. \ref{traj} the  red, green and blue lines represent these three types of trajectory behaviour, respectively.

For $1.32\le c_1\le5$, all the trajectories exit  through the region of the  lower left saddle  of the Caldera PES. This is the phenomenon of dynamical matching \cite{geng2021influence}. In  panel B of Fig. \ref{traj} the red line represent this behavior.

\begin{figure}
 \centering
\includegraphics[scale=0.17]{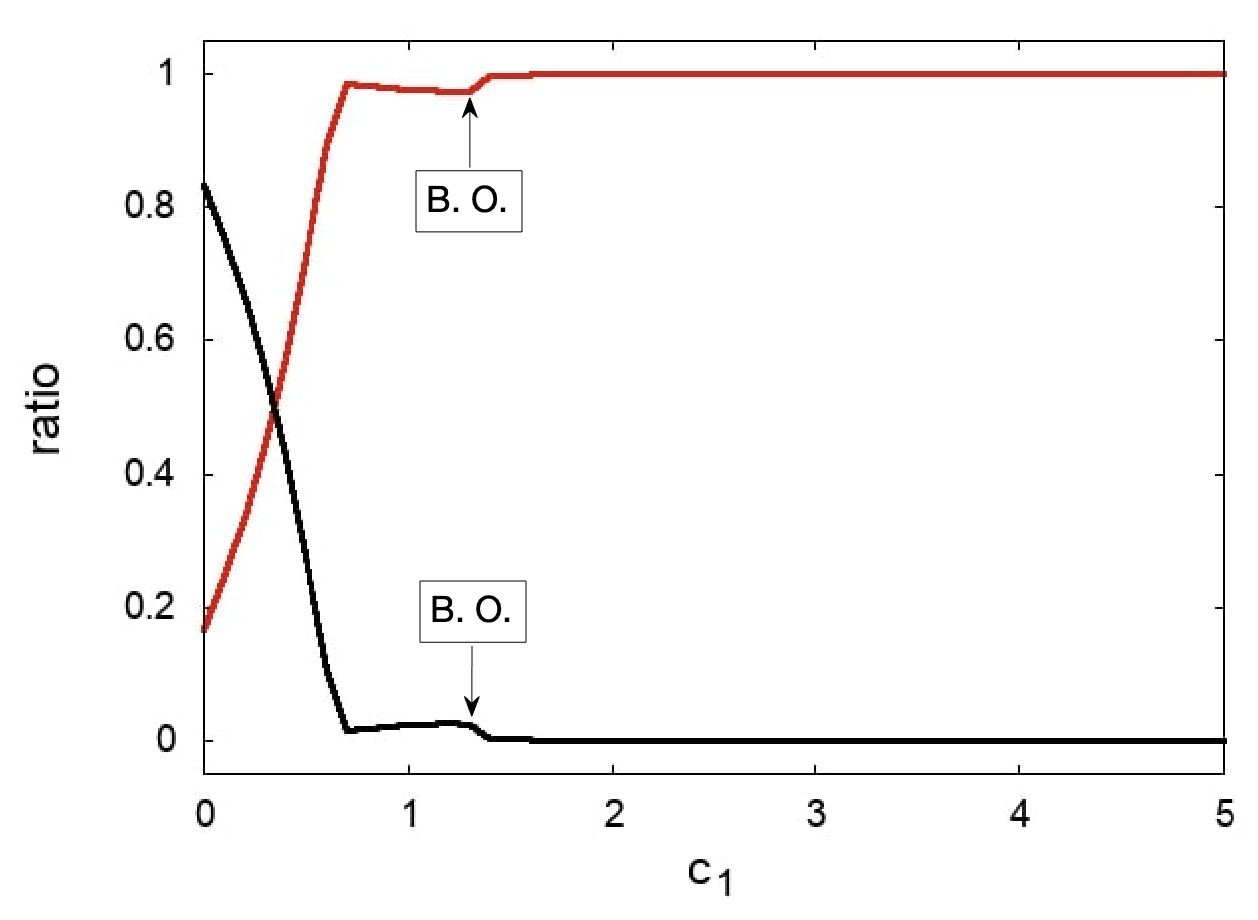}\\
\caption{The ratio of the trajectories that exit through the lower left exit region (or the region of the lower left saddle - red line) and the ratio of the trajectories that are trapped or that exit through the lower right exit region (or the region of the lower right saddle - black line) versus parameter $c_{1}$. With the initials B. O. we denote the critical value of the parameter $c_{1}$ where the bifurcation occurs.}
\label{ratio1}
\end{figure}

We computed the ratio of  trajectories that exited through the lower left exit region (or the region of the lower left saddle)  and the ratio of the trajectories that were trapped or that exited through the lower right exit region (or the region of the lower right saddle)  versus $c_1$.  In Fig. \ref{ratio1}, the red line represents the ratio of the trajectories that exited through the lower left exit region (or the region of the lower left saddle) and the black line represents the ratio of the trajectories that were trapped or that exited through the lower right exit region (or the region of the lower right saddle). We see that initially the black line started at approximately $0.8$ and is higher than the red line  at $0.2$, and afterwards the red line increases to almost $1$ while the black line decreases to almost $0$, after which a plateau occurs for both the red and black lines, and after the critical value of the bifurcation $c_1=1.32$ the final increase and decrease happens for the red and black lines, respectively. Finally, the ratio of the trajectories that exit through the lower left exit region (or the region of the lower left saddle) becomes $1$ and the ratio of the trajectories that are trapped or they exit through the lower right exit region (or the region of the lower right saddle) becomes $0$. Due to the symmetry of the Caldera PES, the conclusion is that all the trajectories coming from one of the upper regions pass through the Caldera and exit through the opposite lower region. This is the phenomenon of dynamical matching in the symmetric Caldera (\cite{geng2021influence}).

\subsection{Trajectory behavior before and after the second Bifurcation}
\label{bif2}

In this section we discuss the behaviour of trajectories for $1\le c_2\le162$ with $c_1=5,c_3=-0.3$ fixed, whose initial conditions are fixed in the region of the upper right index-1 saddle. We have shown (in \cite{geng2021influence} and the previous subsection)  that  dynamical matching is related to a bifurcation effect. However, when we are varying the parameter $c_2$ the bifurcation is not related to a change of the trajectory behavior.

The construction of the initial conditions is the same as  has been explained in the previous Subsection \ref{bif1}. The result from the simulation shows that the phenomenon of dynamical matching exists before the critical value $c_{2} = 22.22$ where the pitchfork bifurcation occurs. Note that for values of $c_{2}$ smaller than the critical value of $c_{2}$ the PES has five critical points. Moreover  dynamical matching persists for values larger than the critical value and  up to the value $c_{2} = 153$, where all the trajectories with initial conditions below the above construction pass through the Caldera and evolve  to the lower left region. We would expect that after  the bifurcation, the dynamical matching would not exist. The figures of two examples of this trajectory behavior (dynamical matching) are given in Fig. \ref{traj} where the parameter $c_{2}$ takes the values B) $c_2=3$ and C) $c_{2} = 153$. These values are chosen to represent dynamical matching before and after the bifurcation, respectively.

\begin{figure}
 \centering
\includegraphics[scale=0.55]{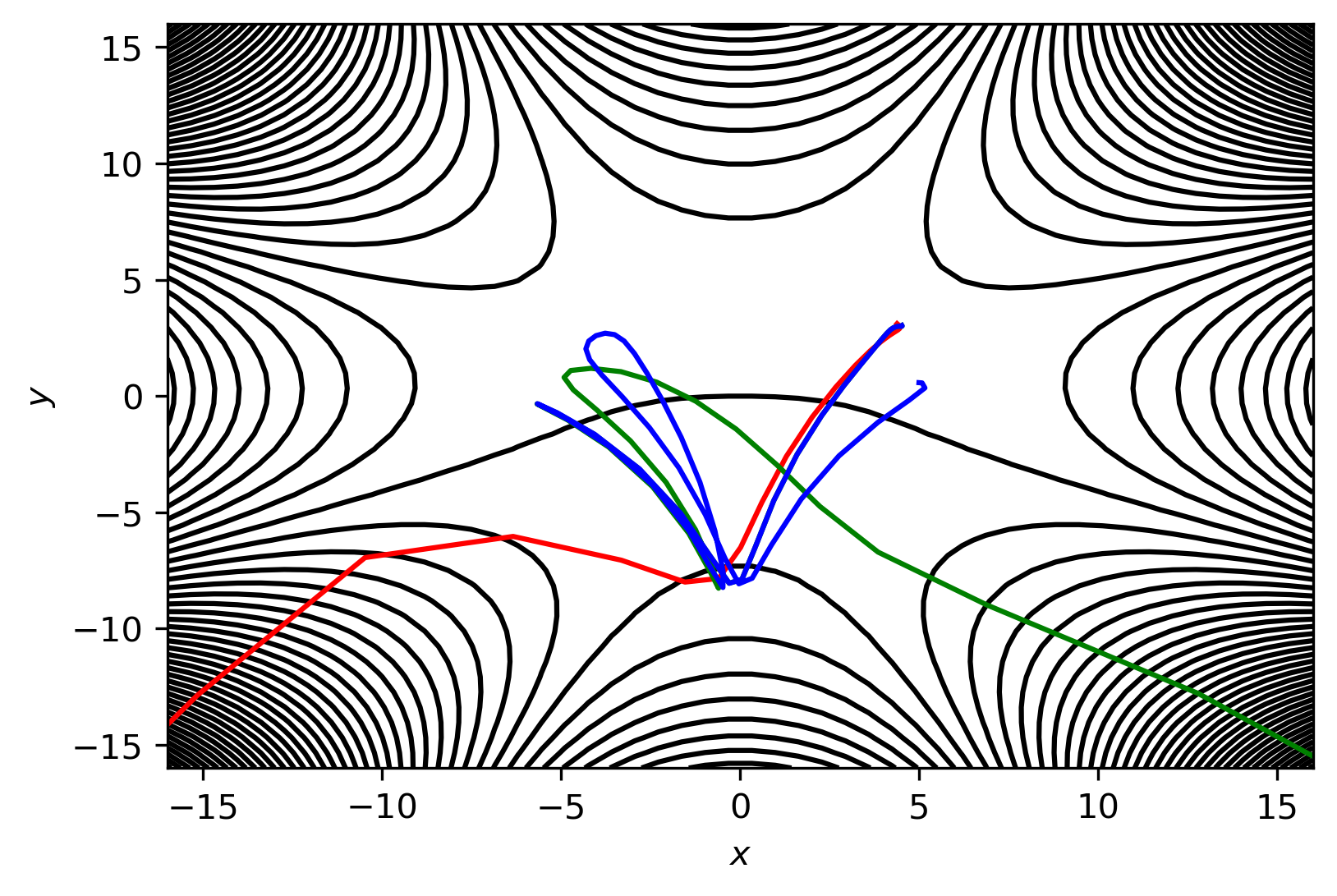}\\
\caption{The contours of the Caldera potential and  trajectories in the configuration space that begin from the region of the upper right saddle (for a value of energy 0.5 units above the energy of the higher index-1 saddles): for $c_1=5,c_2=154$ and $c_3=-0.3$. The trajectories that are trapped or that exit through the lower left exit region  or that exit through the lower right exit region   are depicted by the blue, red and green colors, respectively.}
\label{traj1}
\end{figure}

When $c_2$ is larger than $153$,  dynamical matching is broken and there are three types of trajectory behavior. There are trajectories that exit through the lower left exit region, exit through the lower right exit region, and trajectories that are trapped in the intermediate region of the Caldera. These three types of  trajectory behavior are presented in Fig.  
 \ref{traj1}, where $c_2=154$.

\begin{figure}
 \centering
\includegraphics[scale=0.17]{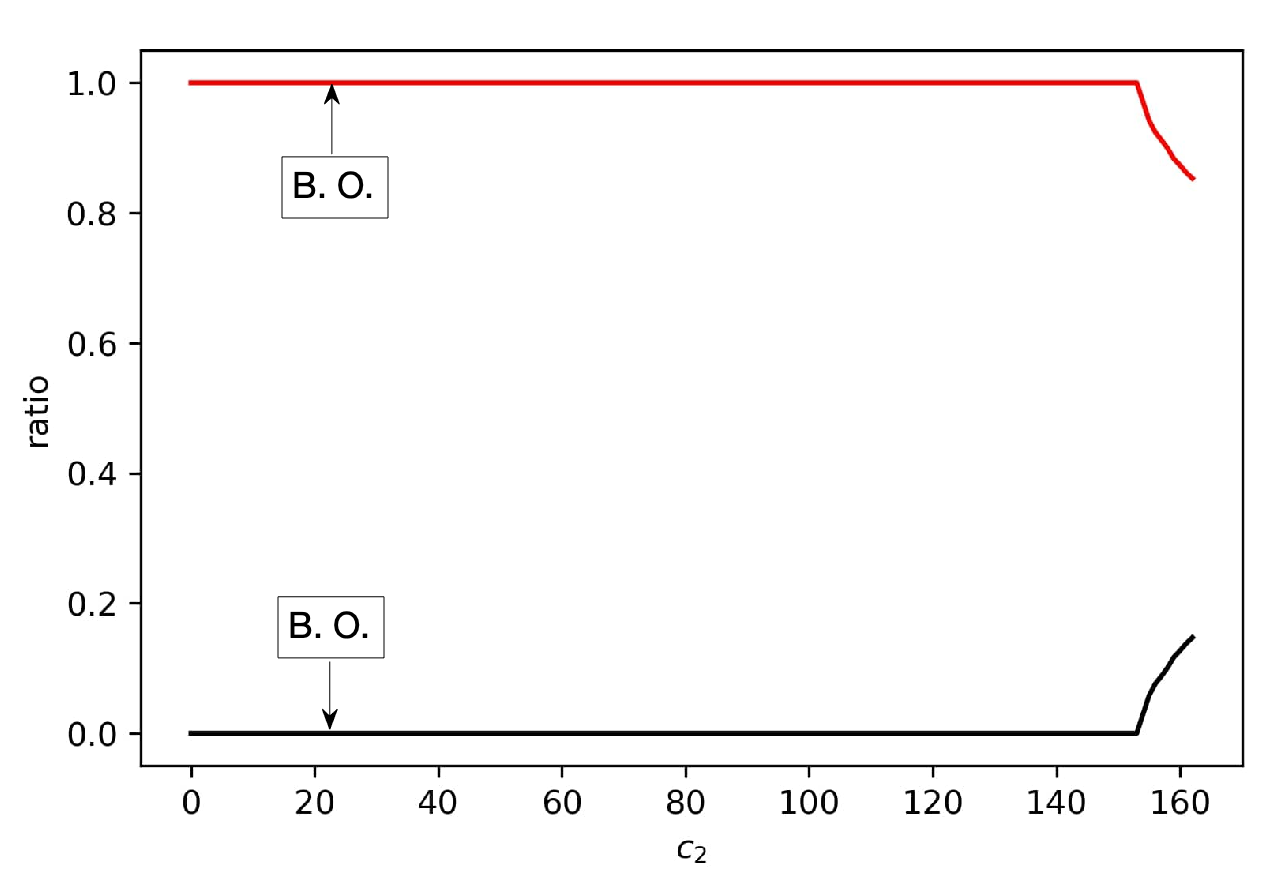}\\
\caption{The ratio of the trajectories that exit through the lower left exit region (or the region of the lower left saddle - red line) and the ratio of the trajectories that are trapped or that exit through the lower right exit region (or the region of the lower right saddle - black line) versus the parameter $c_{2}$. With the initials B. O. we denote the critical value of the parameter $c_{2}$ where the bifurcation occurs. }
\label{ratio2}
\end{figure}

We computed the ratio of the trajectories that exit through the lower left exit region (or the region of the lower left saddle), compared with those trapped in the Caldera or that exit through the lower right exit region (or the region of the lower right saddle), with respect to the parameter $c_2$. We see in Fig. \ref{ratio2} that initially the ratio of the trajectories that exit through the lower left exit region (or the region of the lower left saddle) is $1$ and the ratio of the trajectories trapped in the Caldera or that exit through the lower right exit region (or the region of the lower right saddle) is $0$. This is the phenomenon of dynamical matching in the symmetric Caldera, which is discussed in all previous studies (\cite{collins2014,katsanikas2018,katsanikas2019,katsanikas2020a,katsanikas2020b}). We observed that the bifurcation at $c_2=22.22$ does not result in any change of the ratio, which means that the phenomenon of dynamical matching exists before and after the bifurcation. When $c_2=154$, the ratio of the trajectories that exit through the lower left exit region begins to drop and ends at approximately $0.8$ when $c_2=162$. Conversely, the ratio of the trajectories trapped in the Caldera or that exit through the lower right exit region increases to $0.2$ after $c_2$ reaches $154$. The second bifurcation, that happens at a large value of the depth of the potential, delays this affect on the trajectory behavior of breaking the dynamical matching after the bifurcation point. On the contrary, the first bifurcation (see the Subsection \ref{bif1}),  which happens at a small value of the depth of the potential, affects directly (the change of the trajectory behavior is abrupt) the trajectory behavior after the bifurcation point. This means that the depth of the potential plays a very crucial in the change of the trajectory behavior and whether or not dynamical matching is broken after a bifurcation  of critical points of the symmetric Caldera potential energy surface. 

\section{Conclusions}
\label{conclusions}
In \cite{geng2021influence}  two types of the symmetric Caldera potential energy surface have been considered. The first type has five critical points, one center and four index-1 saddles around it (two upper and two lower saddles). The second type has three critical points, three index-1 saddles (two upper saddles and one lower saddle). In this paper we studied the influence of the depth of the potential on the  trajectory behavior (of trajectories with initial conditions in the region of the upper saddles) before and after  two bifurcations of critical points. These bifurcations represent a transition from  one type of  symmetric Caldera potential to another. The first bifurcation occurs at a small value of the depth of the potential and the second at a large value of the depth of the potential. The main conclusions of our study are the following:

\begin{enumerate}
    \item Dynamical matching is a general feature of the trajectory behavior of the first type of the symmetric Caldera potential energy surface.
    \item We encountered dynamical matching and also the non-existence of dynamical matching as types of  trajectory behavior  of the second type  the  symmetric Caldera potential energy surface.
    \item The depth of the potential is  crucial and controls  the appearance or not of  dynamical matching after a bifurcation of critical points. In the case in which the bifurcation occurs at a small value of the depth of the potential, the change  of the trajectory behavior is abrupt (from  dynamical matching to the non-existence of dynamical matching and vice-versa). On the contrary, we observe a delay of this change of the trajectory behavior if the bifurcation occurs at a large value of the depth of the potential.     
    
\end{enumerate}

\section*{Acknowledgments}
The authors would like to acknowledge the financial support provided by the EPSRC Grant No. EP/P021123/1.

\bibliography{main}

\begin{thebibliography}{10}
\expandafter\ifx\csname url\endcsname\relax
  \def\url#1{\texttt{#1}}\fi
\expandafter\ifx\csname urlprefix\endcsname\relax\def\urlprefix{URL }\fi
\expandafter\ifx\csname href\endcsname\relax
  \def\href#1#2{#2} \def\path#1{#1}\fi

\bibitem{doering1968}
W.~v.~E. Doering, C.~Gilbert, P.~Leermakers, Symmetrical distribution of energy
  in initially unsymmetrically excited products : Reaction of
  dideuteriodiazomethane with allene, methylenecyclopropane and
  vinylcyclopropane, Tetrahedron 24 (1968) 6863--6872.

\bibitem{baldwin2003}
J.~Baldwin, Thermal rearrangements of vinylcyclopropanes to cyclopentenes,
  Chemical Reviews 103~(4) (2003) 1197--1212.

\bibitem{gold1988}
Z.~Goldschmidt, B.~Crammer, Vinylcyclopropane rearrangements, Chem. Soc. Rev 17
  (1988) 229--267.

\bibitem{doubleday1997}
C.~Doubleday, K.~Bolton, W.~Hase, Direct dynamics study of the stereomutation
  of cyclopropane, Journal of the American Chemical Society 119~(22) (1997)
  5251--5252.

\bibitem{doubleday1999}
C.~Doubleday, M.~Nendel, K.~Houk, D.~Thweatt, M.~Page, Direct dynamics
  quasiclassical trajectory study of the stereochemistry of the
  vinylcyclopropane - cyclopentene rearrangement, Journal of the American
  Chemical Society 121~(19) (1999) 4720--4721.

\bibitem{doubleday2006}
C.~Doubleday, C.~Suhrada, K.~Houk, Dynamics of the degenerate rearrangement of
  bicyclo[3.1.0]hex-2-ene, Journal of the American Chemical Society 128~(1)
  (2006) 90--94.

\bibitem{reyes2002}
M.~Reyes, E.~Lobkovsky, B.~Carpenter, Interplay of orbital symmetry and
  nonstatistical dynamics in the thermal rearrangements of
  bicyclo[n.1.0]polyenes, Journal of the American Chemical Society 124 (2002)
  641--651.

\bibitem{collins2014}
P.~Collins, Z.~Kramer, B.~Carpenter, G.~Ezra, S.~Wiggins, Nonstatistical
  dynamics on the caldera, Journal of Chemical Physics 141~(034111) (2014).

\bibitem{katsanikas2018}
M.~Katsanikas, S.~Wiggins, Phase space structure and transport in a caldera
  potential energy surface, International Journal of Bifurcation and Chaos
  28~(13) (2018) 1830042.

\bibitem{geng2021influence}
Y.~Geng, M.~Katsanikas, M.~Agaoglou, S.~Wiggins, The influence of a pitchfork
  bifurcation of the critical points of a symmetric caldera potential energy
  surface on dynamical matching, Chem. Phys. Letters 768 (2021) 138397.

\bibitem{carpenter1985}
B.~K. Carpenter, Trajectories through an intermediate at a fourfold branch
  point. implications for the stereochemistry of biradical reactions, Journal
  of the American Chemical Society 107~(20) (1985) 5730--5732.
\newblock \href {https://doi.org/10.1021/ja00306a021}
  {\path{doi:10.1021/ja00306a021}}.

\bibitem{carpenter1995}
B.~K. Carpenter, Dynamic matching: The cause of inversion of configuration in
  the [1,3] sigmatropic migration?, Journal of the American Chemical Society
  117~(23) (1995) 6336--6344.
\newblock \href {https://doi.org/10.1021/ja00128a024}
  {\path{doi:10.1021/ja00128a024}}.

\bibitem{katsanikas2020a}
M.~Katsanikas, V.~J. Garc\'{i}a-Garrido, S.~Wiggins, The dynamical matching
  mechanism in phase space for caldera-type potential energy surfaces, Chemical
  Physics Letters 743 (2020) 137199.
\newblock \href {https://doi.org/https://doi.org/10.1016/j.cplett.2020.137199}
  {\path{doi:https://doi.org/10.1016/j.cplett.2020.137199}}.

\bibitem{katsanikas2020b}
M.~Katsanikas, V.~J. Garc\'{i}a-Garrido, S.~Wiggins, Detection of dynamical
  matching in a caldera hamiltonian system using lagrangian descriptors, Int.
  J. Bifurcation Chaos 30 (2020) 2030026.

\bibitem{katsanikas2019}
M.~Katsanikas, S.~Wiggins, Phase space analysis of the nonexistence of
  dynamical matching in a stretched caldera potential energy surface,
  International Journal of Bifurcation and Chaos 29~(04) (2019) 1950057.

\end{thebibliography}

\end{document}